\documentclass[prl, aps, twocolumn, amsmath, amssymb, graphicx,superscriptaddress]{revtex4-2}

\usepackage{bm}
\usepackage{graphicx}
\usepackage{braket}
\usepackage[normalem]{ulem}

\usepackage{color}
\usepackage[usenames, dvipsnames]{xcolor}
\usepackage[colorlinks=true]{hyperref}
\hypersetup{linktoc=page, colorlinks=true, linkcolor=MidnightBlue, citecolor=MidnightBlue, urlcolor=OliveGreen}

\usepackage[]{cleveref}
\Crefname{section}{Sec.}{Secs.}
\Crefname{equation}{Eq.}{Eqs.}
\Crefname{figure}{Fig.}{Figs.}
\Crefname{tabular}{Tab.}{Tabs.}

\usepackage{siunitx}
\usepackage{soul}
\usepackage{url}

\usepackage{tabularx}
\usepackage{array}
\newcolumntype{I}{>{\hsize=0.36\hsize}X}
\newcolumntype{C}{>{\hsize=0.3\hsize\centering\arraybackslash}X}
\newcolumntype{U}{>{\hsize=0.35\hsize\centering\arraybackslash}X}
\newcolumntype{V}{>{\hsize=0.4\hsize\centering\arraybackslash}X}
\newcolumntype{W}{>{\hsize=0.44\hsize\centering\arraybackslash}X}
\newcolumntype{Y}{>{\hsize=0.49\hsize\centering\arraybackslash}X}

\usepackage{multirow}
\usepackage{makecell}

\usepackage{comment}

\usepackage[dvipsnames]{xcolor}

\makeatletter
\renewcommand*{\p@subsection}{}
\renewcommand*{\p@subsubsection}{}
\makeatother

\begin{document}

\title{Comment on Hess et al. Phys. Rev. Lett. {\bf 130}, 207001 (2023)}

\author{A. Antipov}
\affiliation{Microsoft Quantum, Station Q, Santa Barbara, CA 93111}
\author{W. Cole}
\affiliation{Microsoft Quantum, Redmond, WA 98052}
\author{K. Kalashnikov}
\affiliation{Microsoft Quantum, Redmond, WA 98052}
\author{F. Karimi}
\affiliation{Microsoft Quantum, Station Q, Santa Barbara, CA 93111}
\author{R. Lutchyn}
\affiliation{Microsoft Quantum, Station Q, Santa Barbara, CA 93111}
\author{C. Nayak}
\affiliation{Microsoft Quantum, Station Q, Santa Barbara, CA 93111}
\author{D. Pikulin}
\affiliation{Microsoft Quantum, Redmond, WA 98052}
\author{G. Winkler}
\affiliation{Microsoft Quantum, Station Q, Santa Barbara, CA 93111}

\date{\today{}}

\begin{abstract}

\end{abstract}

\maketitle

The topological gap protocol (TGP)~\cite{Pikulin21,Aghaee23} gives a binary answer to a binary question:
is there a topological phase present in the (real or simulated) device that produced this transport data set? The TGP was motivated by the fact that binary questions such as this cannot be adequately addressed by qualitative approaches
such as visually comparing conductance spectra.
In this comment, we show that the model introduced in Ref.~\onlinecite{Hess23}
fails the TGP.
In addition, we discuss this model in the broader context of how the TGP
has been benchmarked.

The TGP is designed to identify topological regions in the phase diagram of a device using a measurement and analysis protocol, as detailed in Refs.~\onlinecite{Pikulin21,Aghaee23}.
Note that the TGP does not directly
calculate the topological invariant
(which is not accessible in
transport data) and
serves as a practical experimental proxy for it.
Its accuracy was tested in simulations
in which we can separately compute the topological
invariant and compare it with the output
of the TGP. (All of the data from these simulations and also from our experimental
measurements is available in Ref.~\onlinecite{code_and_data}.)
The TGP will have a low false discovery rate (FDR) ``provided that
the simulated data [used to test the TGP] is drawn from the same probability distribution as the data produced by real devices.''\cite{Aghaee23}
It is possible to design a Hamiltonian that will give a false positive TGP result; its contribution
to the false discovery rate will be weighted by the probability that it occurs in a realistic disorder model.
We characterized disorder in our devices (see Fig. 6 in Ref.~\onlinecite{Aghaee23}) and established that our disorder model is realistic and consistent with the measurements. In other words, we checked our ability to sample in simulations from the same disorder distribution as is present in our physical devices. 
The model in Ref.~\onlinecite{Hess23} is based on specific theoretical configurations of Andreev states that have vanishing probability
of occurring in realistic devices.
Therefore, it could not contribute to the estimated FDR.

\begin{figure}[b!]
\includegraphics[width=8.5cm]{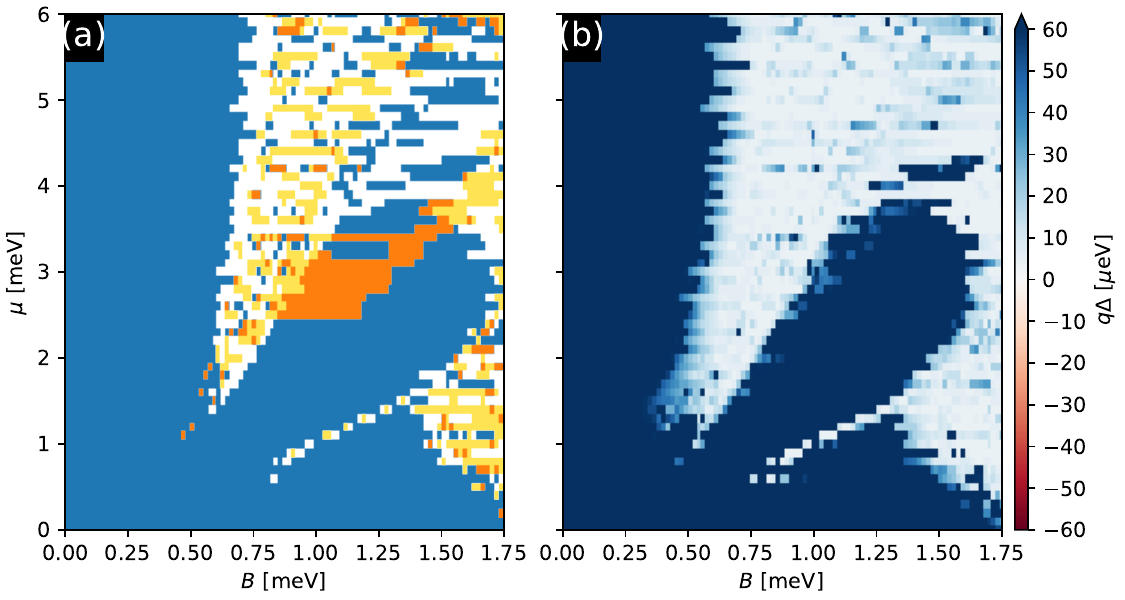}
\caption{
The TGP phase diagram of an exotic Hamiltonian with periodically-modulated Zeeman energy and induced superconductivity \cite{Hess23}. (a) shows the different regions as classified by the TGP and (b) shows the extracted gap $\Delta$ times the topological invariant $q$ as identified by the TGP.
The TGP does not find a topological
region at $\mu=2$, where the simulated conductances show an
accidental gap closing and trivial ZBPs. There are
more promising regions at higher $\mu$ values which
were not discussed \cite{Hess23}, but they do not
pass the TGP either.
}
\vskip 1 cm
\label{fig:Hess-phase-diagram-clean}
\end{figure}

Furthermore, the model of Ref.~\onlinecite{Hess23} fails the TGP. We have made the TGP available \cite{code_and_data} to enable researchers to analyze their experimental data and theoretical models with it. Since this analysis was not carried out in Ref.~\onlinecite{Hess23}, we report the outcome here: it does not pass the TGP. The model's ZBPs and gap closings are insufficient to satisfy all of the TGP's requirements.
Since the model relies on fine-tuning to produce ZBPs on either side (as opposed to a quasi-Majorana scenario that can be more stable~\cite{Prada12, Kells12, Tewari14, LiuC17, Vuik19, Pan21b}) the ZBPs are not stable to variations in the junction region, i.e. changing $\mu_L$ or $\mu_R$ by a little will split the ZBPs. The TGP demands stability to changes of gate voltages that affect the junction region, so the ZBPs in this model would be identified as trivial ZBPs.
If we ignore this for the moment and consider only
one junction configuration then, for the parameter values in Fig. 3 of Ref.~\onlinecite{Hess23} and $T=40$ mK,
we find the TGP phase diagram shown in \Cref{fig:Hess-phase-diagram-clean}.
For the chemical potential value $\mu = 2$ meV  that is plotted in Ref.~\onlinecite{Hess23} as a putative example of a false positive, the ZBP at the right junction is split, so the TGP identifies this region as trivial. 
There is a more promising region at higher chemical potential with ZBPs at both sides, but less than $50\%$ of its boundary is gapless, so it doesn’t pass the TGP there either. If we now vary the junction settings over
the set $\mu_{L,R}=0, \pm0.05, \pm0.1$ meV,
the ZBPs disappear and the orange region in \Cref{fig:Hess-phase-diagram-clean}(a) becomes blue.
(The conductance data and analysis
are available in Ref.~\onlinecite{code_and_data}.)
A different family of Hamiltonians could be designed to produce false positive TGP results. However, as we emphasized above, in order to shed light on the protocol's FDR, we must also determine the probability that it occurs for a realistic disorder distribution.
At the 95\% confidence level,
the results of Ref.~\onlinecite{Aghaee23} indicate that this probability will be less than 8\%.

\bibliographystyle{apsrev}
\bibliography{comment-refs}

\begin{thebibliography}{10}
\expandafter\ifx\csname natexlab\endcsname\relax\def\natexlab#1{#1}\fi
\expandafter\ifx\csname bibnamefont\endcsname\relax
  \def\bibnamefont#1{#1}\fi
\expandafter\ifx\csname bibfnamefont\endcsname\relax
  \def\bibfnamefont#1{#1}\fi
\expandafter\ifx\csname citenamefont\endcsname\relax
  \def\citenamefont#1{#1}\fi
\expandafter\ifx\csname url\endcsname\relax
  \def\url#1{\texttt{#1}}\fi
\expandafter\ifx\csname urlprefix\endcsname\relax\def\urlprefix{URL }\fi
\providecommand{\bibinfo}[2]{#2}
\providecommand{\eprint}[2][]{\url{#2}}

\bibitem[{\citenamefont{Hess et~al.}(2023)\citenamefont{Hess, Legg, Loss, and
  Klinovaja}}]{Hess23}
\bibinfo{author}{\bibfnamefont{R.}~\bibnamefont{Hess}},
  \bibinfo{author}{\bibfnamefont{H.~F.} \bibnamefont{Legg}},
  \bibinfo{author}{\bibfnamefont{D.}~\bibnamefont{Loss}}, \bibnamefont{and}
  \bibinfo{author}{\bibfnamefont{J.}~\bibnamefont{Klinovaja}},
  \bibinfo{journal}{Phys. Rev. Lett.} \textbf{\bibinfo{volume}{130}},
  \bibinfo{pages}{207001} (\bibinfo{year}{2023}).

\bibitem[{\citenamefont{Pikulin et~al.}(2021)\citenamefont{Pikulin, van Heck,
  Karzig, Martinez, Nijholt, Laeven, Winkler, Watson, Heedt, Temurhan
  et~al.}}]{Pikulin21}
\bibinfo{author}{\bibfnamefont{D.~I.} \bibnamefont{Pikulin}},
  \bibinfo{author}{\bibfnamefont{B.}~\bibnamefont{van Heck}},
  \bibinfo{author}{\bibfnamefont{T.}~\bibnamefont{Karzig}},
  \bibinfo{author}{\bibfnamefont{E.~A.} \bibnamefont{Martinez}},
  \bibinfo{author}{\bibfnamefont{B.}~\bibnamefont{Nijholt}},
  \bibinfo{author}{\bibfnamefont{T.}~\bibnamefont{Laeven}},
  \bibinfo{author}{\bibfnamefont{G.~W.} \bibnamefont{Winkler}},
  \bibinfo{author}{\bibfnamefont{J.~D.} \bibnamefont{Watson}},
  \bibinfo{author}{\bibfnamefont{S.}~\bibnamefont{Heedt}},
  \bibinfo{author}{\bibfnamefont{M.}~\bibnamefont{Temurhan}},
  \bibnamefont{et~al.}, \emph{\bibinfo{title}{Protocol to identify a
  topological superconducting phase in a three-terminal device}}
  (\bibinfo{year}{2021}), \eprint{2103.12217}.

\bibitem[{\citenamefont{Aghaee et~al.}(2023)\citenamefont{Aghaee, Akkala, Alam,
  Ali, Alcaraz~Ramirez, Andrzejczuk, Antipov, Aseev, Astafev, Bauer
  et~al.}}]{Aghaee23}
\bibinfo{author}{\bibfnamefont{M.}~\bibnamefont{Aghaee}},
  \bibinfo{author}{\bibfnamefont{A.}~\bibnamefont{Akkala}},
  \bibinfo{author}{\bibfnamefont{Z.}~\bibnamefont{Alam}},
  \bibinfo{author}{\bibfnamefont{R.}~\bibnamefont{Ali}},
  \bibinfo{author}{\bibfnamefont{A.}~\bibnamefont{Alcaraz~Ramirez}},
  \bibinfo{author}{\bibfnamefont{M.}~\bibnamefont{Andrzejczuk}},
  \bibinfo{author}{\bibfnamefont{A.~E.} \bibnamefont{Antipov}},
  \bibinfo{author}{\bibfnamefont{P.}~\bibnamefont{Aseev}},
  \bibinfo{author}{\bibfnamefont{M.}~\bibnamefont{Astafev}},
  \bibinfo{author}{\bibfnamefont{B.}~\bibnamefont{Bauer}}, \bibnamefont{et~al.}
  (\bibinfo{collaboration}{Microsoft Quantum}), \bibinfo{journal}{Phys. Rev. B}
  \textbf{\bibinfo{volume}{107}}, \bibinfo{pages}{245423}
  (\bibinfo{year}{2023}).

\bibitem[{cod()}]{code_and_data}
\bibinfo{note}{The relevant source code for the TGP and the full experimental
and simulated data sets from the paper are available at
  \href{https://github.com/microsoft/azure-quantum-tgp}{github.com/microsoft/azure-quantum-tgp}.}

\bibitem[{\citenamefont{Prada et~al.}(2012)\citenamefont{Prada, San-Jose, and
  Aguado}}]{Prada12}
\bibinfo{author}{\bibfnamefont{E.}~\bibnamefont{Prada}},
  \bibinfo{author}{\bibfnamefont{P.}~\bibnamefont{San-Jose}}, \bibnamefont{and}
  \bibinfo{author}{\bibfnamefont{R.}~\bibnamefont{Aguado}},
  \bibinfo{journal}{Phys. Rev. B} \textbf{\bibinfo{volume}{86}},
  \bibinfo{pages}{180503} (\bibinfo{year}{2012}), \eprint{1203.4488}.

\bibitem[{\citenamefont{Kells et~al.}(2012)\citenamefont{Kells, Meidan, and
  Brouwer}}]{Kells12}
\bibinfo{author}{\bibfnamefont{G.}~\bibnamefont{Kells}},
  \bibinfo{author}{\bibfnamefont{D.}~\bibnamefont{Meidan}}, \bibnamefont{and}
  \bibinfo{author}{\bibfnamefont{P.~W.} \bibnamefont{Brouwer}},
  \bibinfo{journal}{Phys. Rev. B} \textbf{\bibinfo{volume}{86}},
  \bibinfo{pages}{100503} (\bibinfo{year}{2012}).

\bibitem[{\citenamefont{Stanescu and Tewari}(2014)}]{Tewari14}
\bibinfo{author}{\bibfnamefont{T.~D.} \bibnamefont{Stanescu}} \bibnamefont{and}
  \bibinfo{author}{\bibfnamefont{S.}~\bibnamefont{Tewari}},
  \bibinfo{journal}{Phys. Rev. B} \textbf{\bibinfo{volume}{89}},
  \bibinfo{pages}{220507} (\bibinfo{year}{2014}).

\bibitem[{\citenamefont{Liu et~al.}(2017)\citenamefont{Liu, Sau, Stanescu, and
  Das~Sarma}}]{LiuC17}
\bibinfo{author}{\bibfnamefont{C.-X.} \bibnamefont{Liu}},
  \bibinfo{author}{\bibfnamefont{J.~D.} \bibnamefont{Sau}},
  \bibinfo{author}{\bibfnamefont{T.~D.} \bibnamefont{Stanescu}},
  \bibnamefont{and}
  \bibinfo{author}{\bibfnamefont{S.}~\bibnamefont{Das~Sarma}},
  \bibinfo{journal}{Phys. Rev. B} \textbf{\bibinfo{volume}{96}},
  \bibinfo{pages}{075161} (\bibinfo{year}{2017}).

\bibitem[{\citenamefont{Vuik et~al.}(2019)\citenamefont{Vuik, Nijholt,
  Akhmerov, and Wimmer}}]{Vuik19}
\bibinfo{author}{\bibfnamefont{A.}~\bibnamefont{Vuik}},
  \bibinfo{author}{\bibfnamefont{B.}~\bibnamefont{Nijholt}},
  \bibinfo{author}{\bibfnamefont{A.~R.} \bibnamefont{Akhmerov}},
  \bibnamefont{and} \bibinfo{author}{\bibfnamefont{M.}~\bibnamefont{Wimmer}},
  \bibinfo{journal}{SciPost Phys.} \textbf{\bibinfo{volume}{7}},
  \bibinfo{pages}{061} (\bibinfo{year}{2019}).

\bibitem[{\citenamefont{Pan et~al.}(2021)\citenamefont{Pan, Liu, Wimmer, and
  Das~Sarma}}]{Pan21b}
\bibinfo{author}{\bibfnamefont{H.}~\bibnamefont{Pan}},
  \bibinfo{author}{\bibfnamefont{C.-X.} \bibnamefont{Liu}},
  \bibinfo{author}{\bibfnamefont{M.}~\bibnamefont{Wimmer}}, \bibnamefont{and}
  \bibinfo{author}{\bibfnamefont{S.}~\bibnamefont{Das~Sarma}},
  \bibinfo{journal}{Phys. Rev. B} \textbf{\bibinfo{volume}{103}},
  \bibinfo{pages}{214502} (\bibinfo{year}{2021}).

\end{thebibliography}

\end{document}